# Striking Zn impurity effect on the Fe-based superconductor BaFe$_{1.87}$Co$_{0.13}$As$_2$

J. J. Li,[1,2] Y. F. Guo,[3] S. B. Zhang,[3] S. Yu,[1] Y. Tsujimoto,[3] K. Yamaura,[1,2,4,*]

E. Takayama-Muromachi[2,3,4]

[1] Superconducting Materials Center, National Institute for Materials Science, 1-1 Namiki, Tsukuba, Ibaraki 305-0044, Japan

[2] Department of Chemistry, Graduate School of Science, Hokkaido University, Sapporo, Hokkaido 060-0810, Japan

[3] International Center for Materials Nanoarchitectonics (MANA), National Institute for Materials Science, 1-1 Namiki, Tsukuba, Ibaraki 305-0044, Japan

[4] JST, Transformative Research-Project on Iron Pnictides (TRIP), 1-1 Namiki, Tsukuba, Ibaraki 305-0044, Japan

Nonmagnetic impurity effect was studied on the n-type Fe-based superconductor BaFe$_{1.87}$Co$_{0.13}$As$_2$ ($T_c$ = 25 K) by a successful Zn substitution for Fe up to 7 at.%. Magnetic susceptibility, electrical resistivity, specific heat, and Hall coefficient measurements indicated that $T_c$ linearly decreases with the Zn concentration and disappears at 7 at.%. The result is quantitatively comparable with what was observed for YBCO, while it disagrees with a recent report for the p-type Ba$_{0.5}$K$_{0.5}$Fe$_2$As$_2$. Fragile SC against a nonmagnetic impurity was first confirmed for the n-type 122 Fe-based superconductor.

Discovery of the Fe-based superconductor in 2008 attracted increasing attentions due to its fascinating superconductivity (SC).[1-3] The layered structure forms the anisotropic Fe-3$d$ electronic state and the SC appears in the vicinity of the antiferromagnetically ordered state, implying a tight relation between the SC and the magnetism.[4] Right after the discovery many fundamental questions were aroused regarding the pairing symmetry of the SC.[4,5] Early theoretical studies proposed a multi-gaped $s$-wave (often notated as $s_\pm$ wave) model for the pairing,[6-8] and afterwards it was experimentally supported by such as NMR studies,[9,10] angle-resolved photoemission spectroscopy,[11,12] microwave penetration depth measurements,[13] μSR studies,[14,15] and neutron scattering studies.[16,17] Recent half-flux quantum experiments on NdFeAsO$_{0.88}$F$_{0.12}$ also supported the model.[18] Impurity effect studies, however, did not support it because a doped impurity (such as Co, Ni, Ru, Rh, Pd, or Ir) was found even to contribute to establishing the SC in contrast to expectations from the $s_\pm$ wave model.[6-8] Discussion about the paring mechanism was thus complicated.[19-23]

Generally, studying a nonmagnetic impurity (NMI) effect on SC helps greatly to investigate the pairing symmetry. For instance, Anderson's theorem predicts that a NMI cannot break pairing of isotropic superconducting gap, but anisotropic gap in contrast.[24] The predictions well met so-far experiments; MgB$_2$, an almost isotropically gapped superconductor, shows indeed robust SC against doped Zn,[25] while the cuprate superconductor, possessing anisotropic gap, immediately loses the SC even by a small amount of doped Zn.[26-29] NMI studies by Zn were, hence, expected to help to investigate the pairing symmetry of the Fe-based superconductor. However, even an experimental consensus has been controversial; Guo *et al.* found that the doped Zn efficiently suppresses the SC of LaFeAsO$_{0.85}$ in roughly linear as well as that of LaFeAsO$_{0.85}$F$_{0.15}$, while many incompatible results were obtained theretofore.[30,31] Cheng *et al.* reported that the Zn impurity hardly affects the SC of the p-type Ba$_{0.5}$K$_{0.5}$Fe$_2$As$_2$.[32] To establish a solid picture of the NMI effect on the Fe-based superconductor, additional Zn studies are necessary. A comprehensive picture of the NMI effect will promote better understanding of the Fe-based superconductor.

In this study, Zn concentration of the n-type 122 superconductor BaFe$_{1.87}$Co$_{0.13}$As$_2$ was successfully controlled by employing a high-pressure and high-temperature heating method. The magnetic and electrical measurements revealed that the SC is fragile against a doped Zn. These results thus suggest that the Zn study is indeed capable of investigating a NMI effect on the Fe-based superconductor because of the highly localized $d^{10}$ state in the Fe$_2$As$_2$ layer.[33] This paper reports the first observation of the fragile SC against a NMI for the n-type 122 Fe-based superconductor.

Single crystalline-like BaFe$_{1.87-x}$Zn$_x$Co$_{0.13}$As$_2$ ($x$ = 0, 0.02, 0.04, 0.06, 0.08, 0.10, 0.12 and 0.14) were prepared as follows: each mixture of BaAs (lab made), FeAs (lab made), Fe (3N), Co ($\geq$99.5%) and Zn (4N) was placed into a Ta capsule with an h-BN inner (preheated at ~1900 °C for 1 h in nitrogen). The loaded capsule was treated at 3 GPa in a belt-type high-pressure apparatus at 1300 °C for 2 hrs, and the temperature was decreased to 1100 °C for 2 hrs. The capsule was then quenched to room temperature, followed by releasing the pressure. The as-prepared samples were kept in vacuum for 3-5 days, which resulted in isolation of small single-like crystals with sizes up to ~0.5 mm in the largest dimension. Perhaps, this is due to FeAs which works as a flux during the heating.[34] The crystals were crushed into powder and characterized by an X-ray diffraction (XRD) measurement using Cu-K$\alpha$ radiation. Selected crystals with $x$ of 0, 0.04, 0.06, 0.08 and 0.12 were studied in an electron probe micro-analysis (EPMA; JXA-8500F, JEOL).

Magnetic susceptibility ($\chi$) of an amount of small crystals (~30 mg) was measured in Magnetic Properties Measurement System, Quantum Design, from 2 to 35 K in a magnetic field ($H$) of 10 Oe, here the crystals were loosely gathered in a sample holder. Electrical resistivity ($\rho$) of selected crystals was measured between 2 and 300 K in Physical Properties Measurement System (PPMS), Quantum Design, in various $H$ up to 70 kOe. We measured the Hall coefficient ($R_\mathrm{H}$) of selected single-like crystals with $x$ of 0, 0.04 and 0.08. For each sample with the amount of 12-14 mg crystals, we measured the specific heat ($C_\mathrm{p}$) in PPMS from 2 to 300 K by a heat-pulse relaxation method. In the $\rho$ and $R_\mathrm{H}$ measurements we were unable to control the crystal direction because of the crystal size and possible multiple domains.

The powder XRD patterns of BaFe$_{1.87-x}$Zn$_x$Co$_{0.13}$As$_2$ (Fig. 1a) clearly indicate that the tetragonal ThCr$_2$Si$_2$-type structure (*I*4/*mmm*) is formed over the composition range from $x = 0$ to 0.14.[35] The lattice constants estimated are summarized in Fig. 1b. A nearly isotropic expansion of *a* and *c* is observed (see the inset to Fig. 1a; peaks gradually shift over the Zn substitution), reflecting difference between the Zn-As and the Fe-As bonds as discussed in ref. 31. Besides, a magnetic effect is possibly somewhat involved in the c-axis expansion as discussed for LaFe$_{1-x}$Zn$_x$AsO.[33] Zn substitution for Cu in YBa$_2$Cu$_3$O$_{7-\delta}$ also results in an isotropic expansion of the lattice constants.[36]

The Zn substitution was again confirmed in EPMA: the true atomic ratio Zn/Ba was 0 ($x = 0$), 0.043(3) (0.04), 0.066(2) (0.06), 0.088(1) (0.08), and 0.161(9) (0.12), indicating little gap between the nominal and the true Zn concentration. Co/Ba was 0.1133(7) ($x = 0$, Co$_{0.13}$), 0.1092(7) (0.04), 0.1151(8) (0.06), 0.1049(7) (0.08), and 0.1056(8) (0.12), indicating the constant Co concentration over the Zn substitution. Generally, growth of a Zn doped crystal is often unsuccessful because of high volatility of Zn during heating, our results however indicate the Zn substitution is controlled for BaFe$_{1.87}$Co$_{0.13}$As$_2$ up to 7 at.%. Perhaps, the high-pressure condition was effective to suppress the volatility.

Fig. 2a shows *T* dependence of $\chi$ of BaFe$_{1.87-x}$Zn$_x$Co$_{0.13}$As$_2$ ($x = 0$-0.14). The host crystal BaFe$_{1.87}$Co$_{0.13}$As$_2$ shows quite similar superconducting features with those reported elsewhere.[37,38] Obviously, $T_c$ decreases continuously with increasing the Zn content; $T_c$ is 25, 20, 18, 17 and 11 K at *x* of 0, 0.02, 0.04, 0.06 and 0.08, respectively. Note that the SC completely disappears at $x = 0.14$ (7 at.% of Zn). For further characterization, magnetization loops at 2 K for all samples were measured, as shown in Fig. 2b. The loop becomes smaller with increasing the Zn concentration, corresponding to the $T_c$ decreases. Note that the compound at *x* of 0.08 shows a weak ferromagnetic behavior, although the temperature is still below $T_c$ (~11 K). Local magnetic moments around the Zn atom in the superconducting layer are possibly responsible for this, as discussed for the Zn-doped Bi$_2$Sr$_2$CaCu$_2$O$_8$.[39]

The ρ vs. $T$ curves for $BaFe_{1.87-x}Zn_xCo_{0.13}As_2$ between 2 K and 300 K are shown in Fig. 3a. The $T_c$ of the Zn-free crystal is 26.2 K, slightly larger than the magnetically estimated $T_c$. The data clearly indicate that $T_c$ goes down with increasing the Zn concentration, as found in the magnetic study. It is noteworthy that the non-SC sample remains metallic (see the $x = 0.14$ curve for example). The residual ρ (1.27 mΩ-cm) is indeed larger than that of the Zn-free sample (0.26 mΩ-cm), however neither Anderson's localization-like nor Kondo-like interactions contribute to the ρ-$T$ curves. It is thus most likely that the doped Zn effects pair-breaking because of scatterings associated with the highly localized state of Zn in the $Fe_2As_2$ layer, as was discussed in Ref. 33.

Fig. 3b shows ρ curves for $BaFe_{1.87-x}Zn_xCo_{0.13}As_2$ measured in dc $H$ up to 70 kOe. All curves show a sharp transition even in the highest $H$ of 70 kOe, indicating a homogeneously distribution of Zn throughout the crystal. Moreover, the crystals with $x = 0.08$ or below show an obvious "parallel shift" in the curves by a rate of $-0.5\times10^{-4}$ K/Oe. It is interesting that a comparable feature was observed for the Zn-doped $(La,Sr)_2CuO_4$, which Kofu *et al*. attributed to the gradually filled spin gap and appearance of incommensurately static magnetic correlations.[40] In contrast, the highly Zn-doped sample ($x = 0.14$) display neither a $H$-dependent nor a $T$-dependent feature, indicating a normal metallic behavior.

The upper critical field ($H_{c2}$) was estimated from the ρ-$T$ curves (Fig. 3c). Here, $H_{c2}$ is defined as $H_{10}$ or $H_{90}$, as ρ($T$) is equal to 10 % or 90 % of the normal state value, respectively. We determined $H_d$ from a peak in the $d\rho/dT$ vs. $T$ plot (not shown). All the $H_{c2}$-$T$ curves are roughly linear; we thus fit the curves using a linear function $H_{c2}(T) = H_0 + T\partial H_{c2}/\partial T$, where $H_0$ corresponds to $H_{c2}(T=0)$. The slope apparently shows negligibly small difference among the definitions. Using the WHH model $\mu_0 H_{c2}(0) = -0.693(dH_{c2}/dT)_{T=T_c} T_c$ [41] we can roughly estimate $\mu_0 H_{c2}(0)$ of 65, 35, 32, 40, and 21 T at $x = 0$, 0.02, 0.04, 0.06 and 0.08, respectively. The $\mu_0 H_{c2}(0)$ for the Zn-free crystal is close to the value reported elsewhere.[37,38,42]

The superconducting coherence length (ξ) is calculated using the equation

$\xi = \sqrt{\Phi_0 / 2\pi H_{c2}(0)}$, where $\Phi_0$ is the flux quantum.[40,42] We obtained $\xi$ of 22, 30, 32, 28 and 39 Å, respectively. The $\xi$ of the Zn-doped crystals are slightly larger than that of the Zn-free crystal. According to the "Swiss cheese" model proposed by Nachumi *et al.*, charge carriers are excluded from super-fluid within a Zn-centered non-SC island, whose radius is $\xi_{nSC}$.[28] Once the average distance among Zn ions ($l_i$) becomes larger than $\xi_{nSC}$, the neighbor non-SC islands are connected. The incommensurately magnetic correlations thus can be stable, which may account for the parallel shift observed for the $\rho(T)$-$H$ curves, as discussed in Ref. 40. Here, $l_i$ is given by $l_i(x) = s/\sqrt{0.5x}$, where $s$ is the distance between Fe ions (2.77 Å).[35] It should be noted that the $l_i$ (28 Å) roughly accesses to the $\xi$, being comparable with features of the Zn-doped $(La,Sr)_2CuO_4$.[40]

Fig. 4a shows the $T$ dependence of $R_H$ for $BaFe_{1.87-x}Zn_xCo_{0.13}As_2$ with $x$ = 0, 0.04 and 0.08. The data for the Zn-free crystal accesses to the early data.[43] Regarding the normal state, we found there is no significant change over the Zn substitution, indicating that the Zn substitution do not substantially alter the actual carrier density. The estimated carrier density at 50 K is $7.99 \times 10^{20}$ ($x$ = 0), $1.67 \times 10^{21}$ (0.04), and $4.67 \times 10^{20}$ (0.08). Regarding the Zn-doped YBCO, fairly little change was observed in the $R_H$ measurements as well.[36] The present results are reasonable because the substitution is isovalent. Moreover, the compounds are oxygen-free; unavoidable impacts on $T_c$ due to an oxygen nonstoichiometry can be eliminated.

The $C_p$ vs. $T$ and $C_p/T$ vs. $T^2$ plots for the crystals with $x$ = 0, 0.04 and 0.08 are given in Fig. 4b. It appears that an expected anomaly at $T_c$ is fairly small in $C_p$ vs. $T$ even at $x$ = 0. It is possible that undetected disorders broaden the SC peak. Such the weak anomaly was observed for the Co doped single-crystals of $BaFe_2As_2$ by others.[37,38] Note that the $C_p/T$ extrapolated to the zero temperature ($\equiv \gamma_0$) is enhanced by increasing the dopant of Zn as seen in the $C_p/T$ vs. $T^2$ plot in the inset to Fig. 4b; the $\gamma_0$ increases from ~6 mJ mol$^{-1}$ K$^{-2}$ for the Zn-free crystal to ~20 and ~37 mJ mol$^{-1}$ K$^{-2}$ for the $x$ = 0.04 and 0.08 crystals. A comparable enhancement was found not only for the Zn doped YBCO [44] and LSCO [45] but also for the Zn doped $LaFeAsO_{0.85}$.[30] Regarding the Zn doped YBCO, pair-breaking effect of the Zn substitution and Kondo-screened moments were discussed to account for the $\gamma_0$ enhancement, while the Zn-centered non-SC region $\pi(\xi_{nSC})^2$ for LSCO was considered to contribute to the increasing residual density of states at the Fermi level and consequently the enhanced $\gamma_0$. The Zn doped $BaFe_{1.87}Co_{0.13}As_2$ needs to be further investigated to reveal nature of the $\gamma_0$ enhancement in addition to the small jump at 5 K ($x = 0.08$).

Fig. 5 is an experimental summary of the Zn doped studies over the contemporary superconducting materials. All $T_c/T_{c0}$ appear to depend almost linearly on $n$ ($= x/2$), we thus applied a linear fit to the data (solid lines in Fig. 5). From the fits, we estimated $T_c$ decreasing rate for $BaFe_{1.87-2n}Zn_{2n}Co_{0.13}As_2$ to be -14 per $n$, while -24, -23, -12, -42, <-1 were found for $Nd_{1.85}Ce_{0.15}Cu_{1-n}Zn_nO_4$,[46] $La_{1.85}Sr_{0.15}Cu_{1-n}Zn_nO_4$,[46] $YBa_2Cu_{3-3n}Zn_{3n}O_{6.93}$,[27] $LaFeAs_{1-n}Zn_nO_{0.85}$,[30] $Mg_{1-n}Zn_nB_2$,[25] respectively. It appears that the rate for $BaFe_{1.87}Co_{0.13}As_2$ is nearly comparable with that for $YBa_2Cu_3O_{6.93}$ and distinguishably larger than that for $MgB_2$. We thus conclude that the doped impurity Zn suppresses the SC of $BaFe_{1.87}Co_{0.13}As_2$ as efficiently as it does for $YBa_2Cu_3O_{6.93}$. It is noteworthy that the rate for $LaFeAsO_{0.85}$ is three times larger than that for $BaFe_{1.87}Co_{0.13}As_2$, possibly reflecting difference between the actual electronic states of the 1111 and 122 structures.

Nakajima *et al.* very recently reported effects of nonmagnetic point defects introduced by proton irradiation on $Ba(Fe,Co)_2As_2$,[47] showing similar results as those of our chemical doping. Cheng *et al.* founded that the Zn impurity hardly affects the SC of the hole-doped $Ba_{0.5}K_{0.5}Fe_2As_2$.[32] It was suggested that the degree of $T_c$ suppression depends on the doping level.[31] Recent NMR studies on the P-doped $BaFe_2As_2$ [48] and theoretical studies on the local structure of the $Fe_2As_2$ layer [49] indicated that the gap symmetry can possibly change depending on minute factors. A *d*- to *s*-wave change was predicted theoretically to depend on degree of disorder.[50] In order to obtain a comprehensive picture of a NMI effect on the Fe-based superconductor, further Zn studies over varieties of the Fe-based superconductors, including the 11, 111, 122, and 1111 systems, are indispensable from p-doped to n-doped.

In summary, we studied a Zn doping effect on $BaFe_{1.87}Co_{0.13}As_2$. We found that 7 at.% of Zn at most completely suppresses the SC, which is qualitatively comparable with what was observed for YBCO. This paper reports the first observation of the fragile SC against a NMI for the n-type 122 Fe-based superconductor. If Zn scatterings play a primary role of pair-breakings, the result well accords with the prediction from the $s_{\pm}$-wave model,[6-8] while a *d*-wave model remains possible. Besides, the conventional *s*-wave model, suggested by the early impurity effect studies using such as Co, Ni, Ru, Rh, Pd, and Ir, can be excluded. The present transport data and the theoretical studies [33] strongly support the Zn scatterings picture.

We thank M. Miyakawa for the high-pressure experiments and K. Kosuda for the EPMA study. This research was supported in part by the World Premier International Research Center (WPI) Initiative on Materials Nanoarchitectonics from MEXT, Japan; the Grants-in-Aid for Scientific Research (20360012,22246083) from JSPS, Japan; and the Funding Program for World-Leading Innovative R&D on Science and Technology (FIRST Program) from JSPS.

* YAMAURA.Kazunari@nims.go.jp

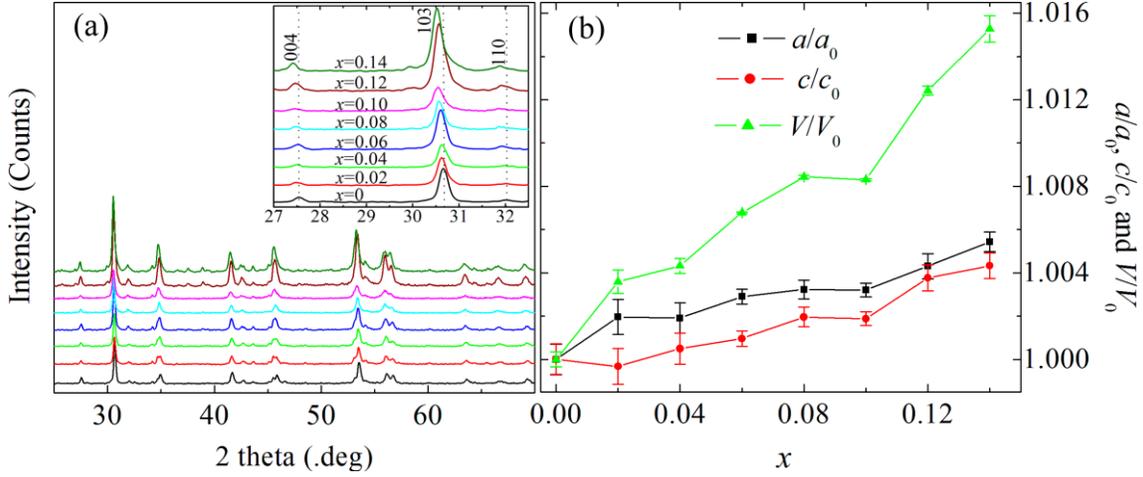

Fig. 1 (a) Powder XRD patterns of BaFe$_{1.87-x}$Zn$_x$Co$_{0.13}$As$_2$ ($x$ = 0-0.14). Inset is an expanded view of the patterns. (b) Lattice parameters vs. nominal Zn-concentration, where $a_0$= 3.955(3) Å, $c_0$= 12.98(1) Å, and $V_0$= 203.0(1) Å$^3$.

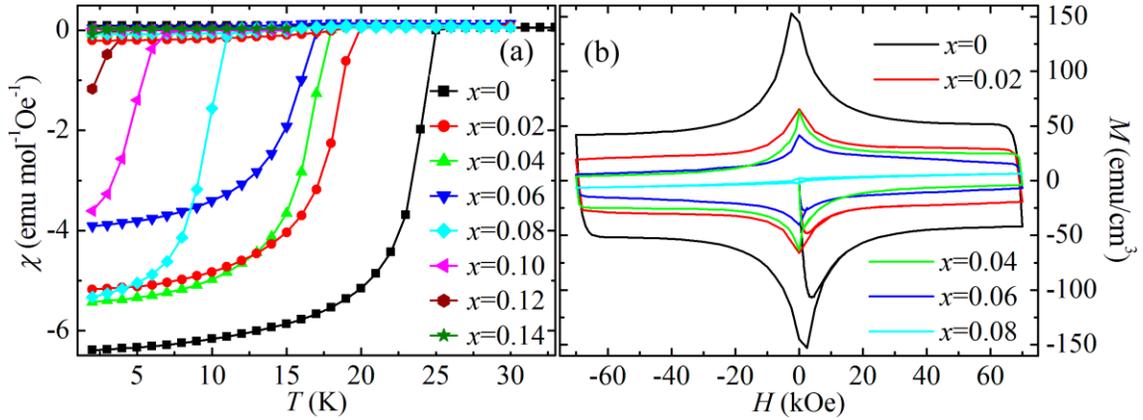

Fig. 2 (a) χ vs. $T$ for BaFe$_{1.87-x}$Zn$_x$Co$_{0.13}$As$_2$ ($x$ = 0-0.14) at $H$ = 10 Oe, and (b) $M$ vs. $H$ at $T$ = 2 K.

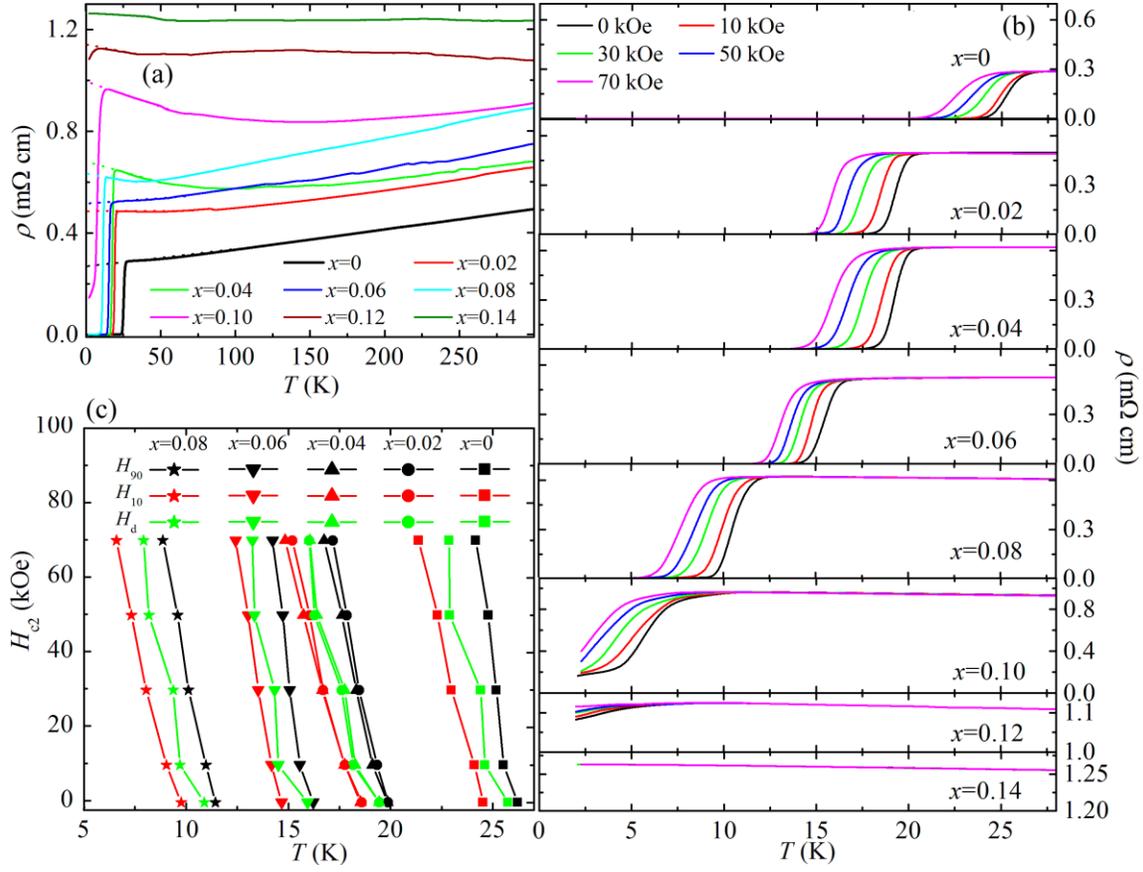

Fig. 3  (a) ρ vs. T for the single-crystal BaFe$_{1.87-x}$Zn$_x$Co$_{0.13}$As$_2$ ($x$ = 0-0.14), and (b) an expanded view with the data measured in $H$ of 10, 30, 50, and 70 kOe. (c) $H_{c2}$ vs. $T$ calculated from the data using various definitions.

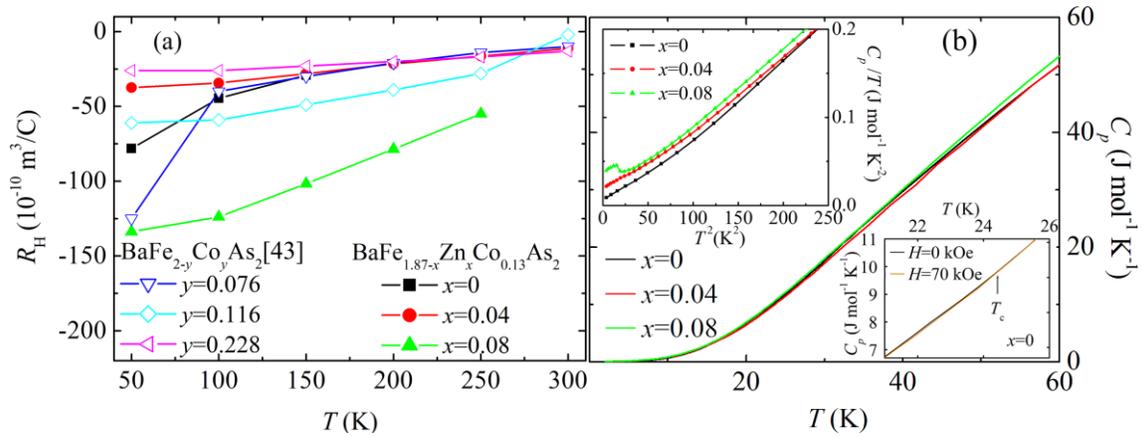

Fig. 4  $T$ dependence of (a) $R_H$ and (b) $C_p$ for BaFe$_{1.87-x}$Zn$_x$Co$_{0.13}$As$_2$ ($x$ = 0, 0.04 and 0.08).

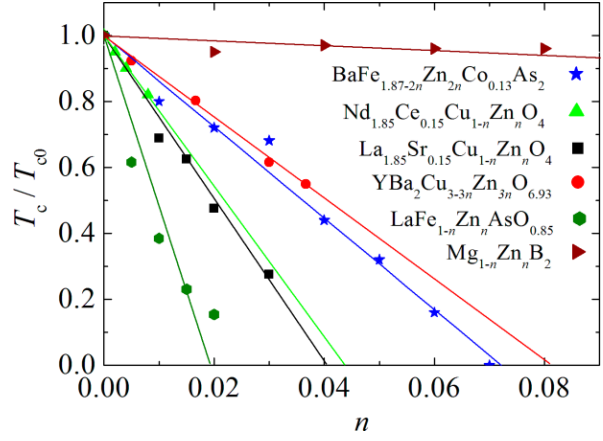

Fig. 5　$n$ dependence of $T_c/T_{c0}$, where $T_{c0}$ corresponds to $T_c$ at $n = 0$, for comparing Zn-substituted superconductors. Solid lines are fits to the data by using a linear function.